\documentclass{vgtc}                          

\ifpdf
  \pdfoutput=1\relax                   
  \usepackage{graphicx}                
  \DeclareGraphicsExtensions{.pdf,.png,.jpg,.jpeg} 
\else
  \usepackage{graphicx}                
  \DeclareGraphicsExtensions{.eps}     
\fi%

\graphicspath{{figures/}{pictures/}{images/}{./}} 

\usepackage{microtype}                 
\PassOptionsToPackage{warn}{textcomp}  
\usepackage{textcomp}                  
\usepackage{mathptmx}                  
\usepackage{times}                     
\usepackage{cite}                      
\usepackage{tabu}                      
\usepackage{booktabs}                  

\usepackage{enumitem}

\usepackage{comment}

\onlineid{0}

\vgtccategory{Research}

\vgtcinsertpkg

\title{Hanstreamer: an Open-source Webcam-based\\
Live Data Presentation System}

\author{Adrian Kristanto\thanks{e-mail: adrian.kristanto@monash.edu}\\ %
        \scriptsize Monash University %
\and Maxime Cordeil\thanks{e-mail: m.cordeil@uq.edu.au}\\ %
     \scriptsize University of Queensland
\and Benjamin Tag\thanks{e-mail: benjamin.tag@monash.edu}\\ %
     \scriptsize Monash University
\and Nathalie Henry Riche\thanks{e-mail: Nathalie.Henry@microsoft.com}\\ %
     \scriptsize Microsoft Research
\and Tim Dwyer\thanks{e-mail: tim.dywer@monash.edu}\\ %
     \scriptsize Monash University
     }

\teaser{
  \includegraphics[width=0.33\textwidth]{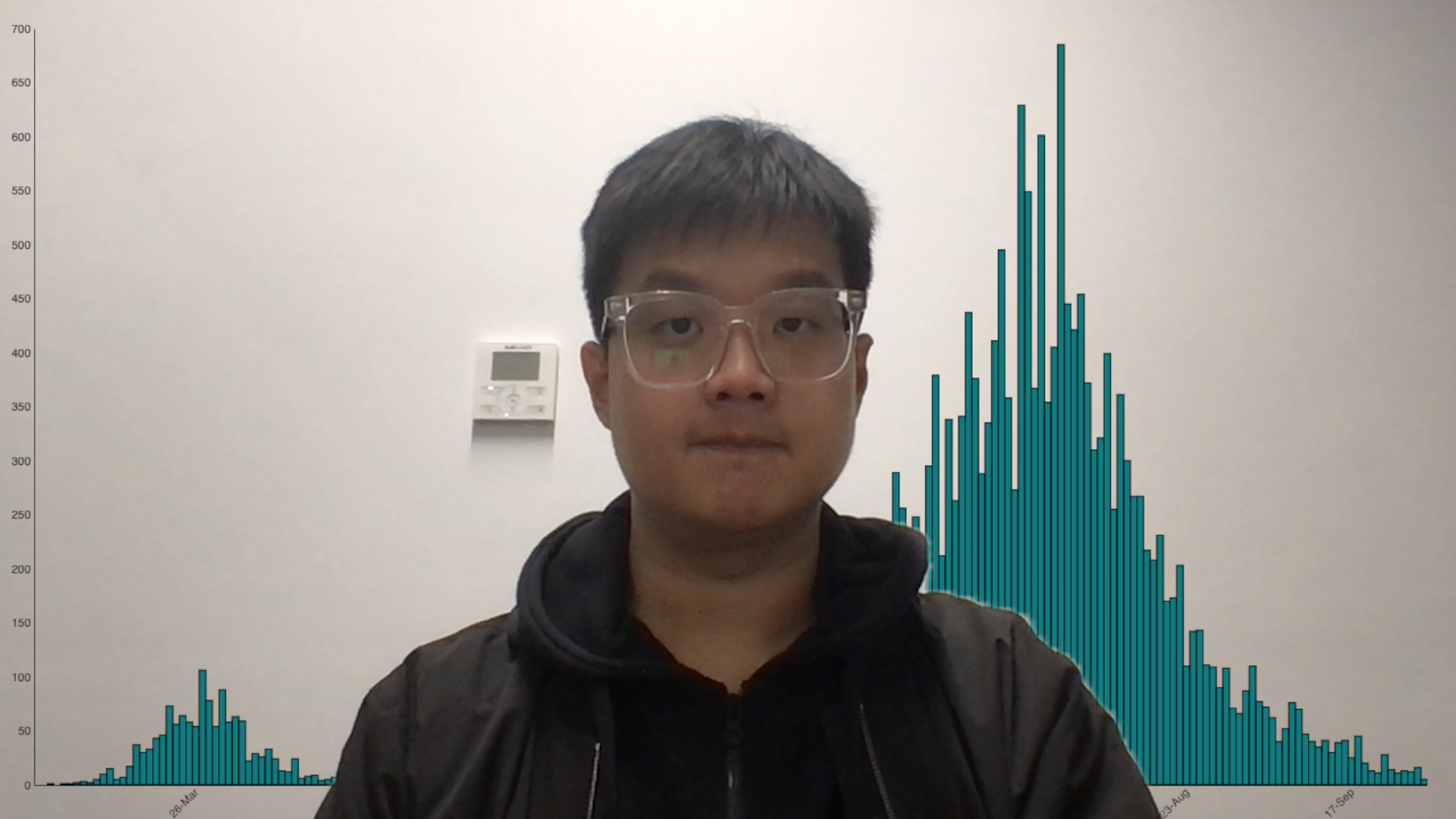}
\includegraphics[width=0.33\textwidth]{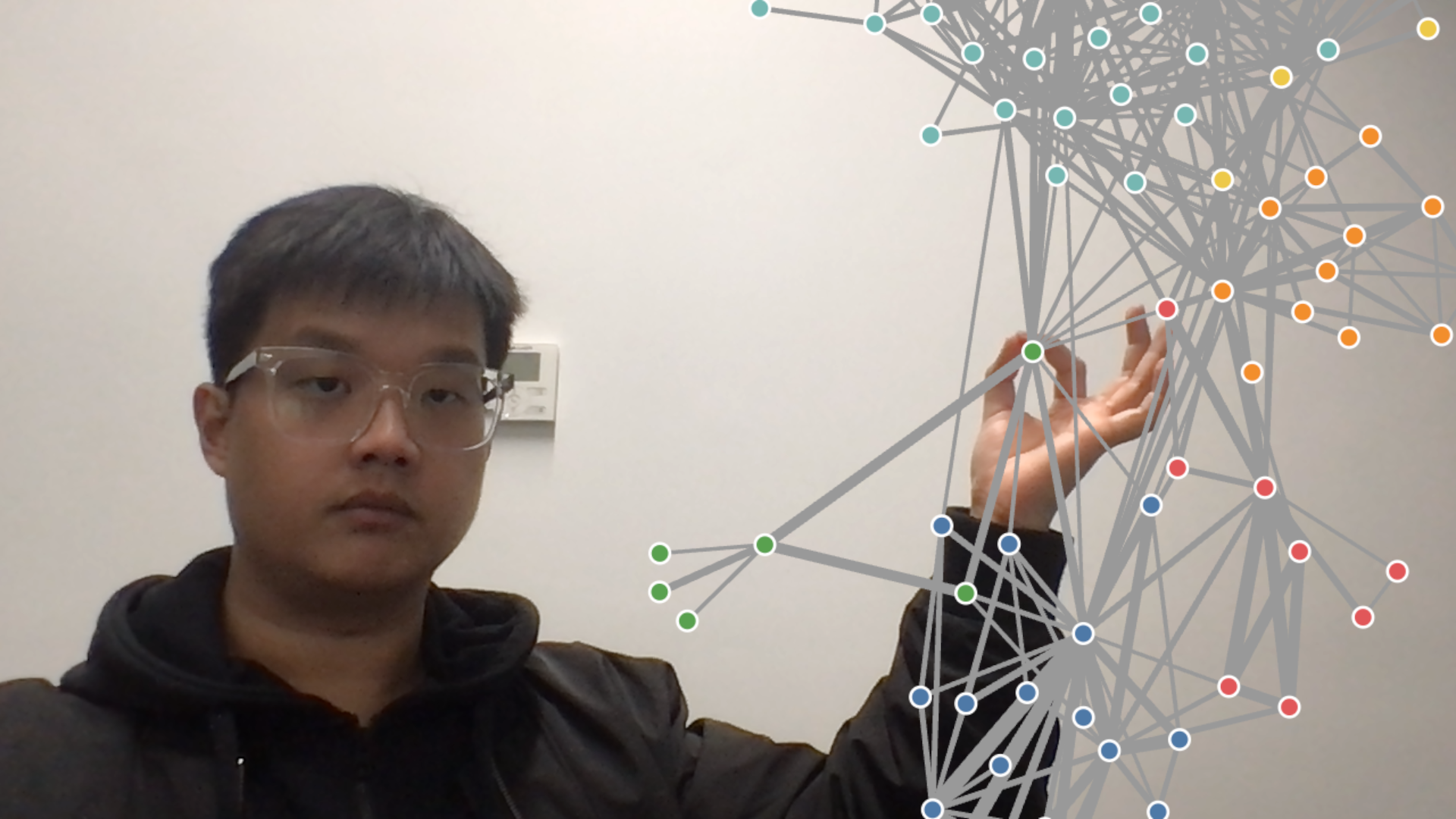}
\includegraphics[width=0.33\textwidth]{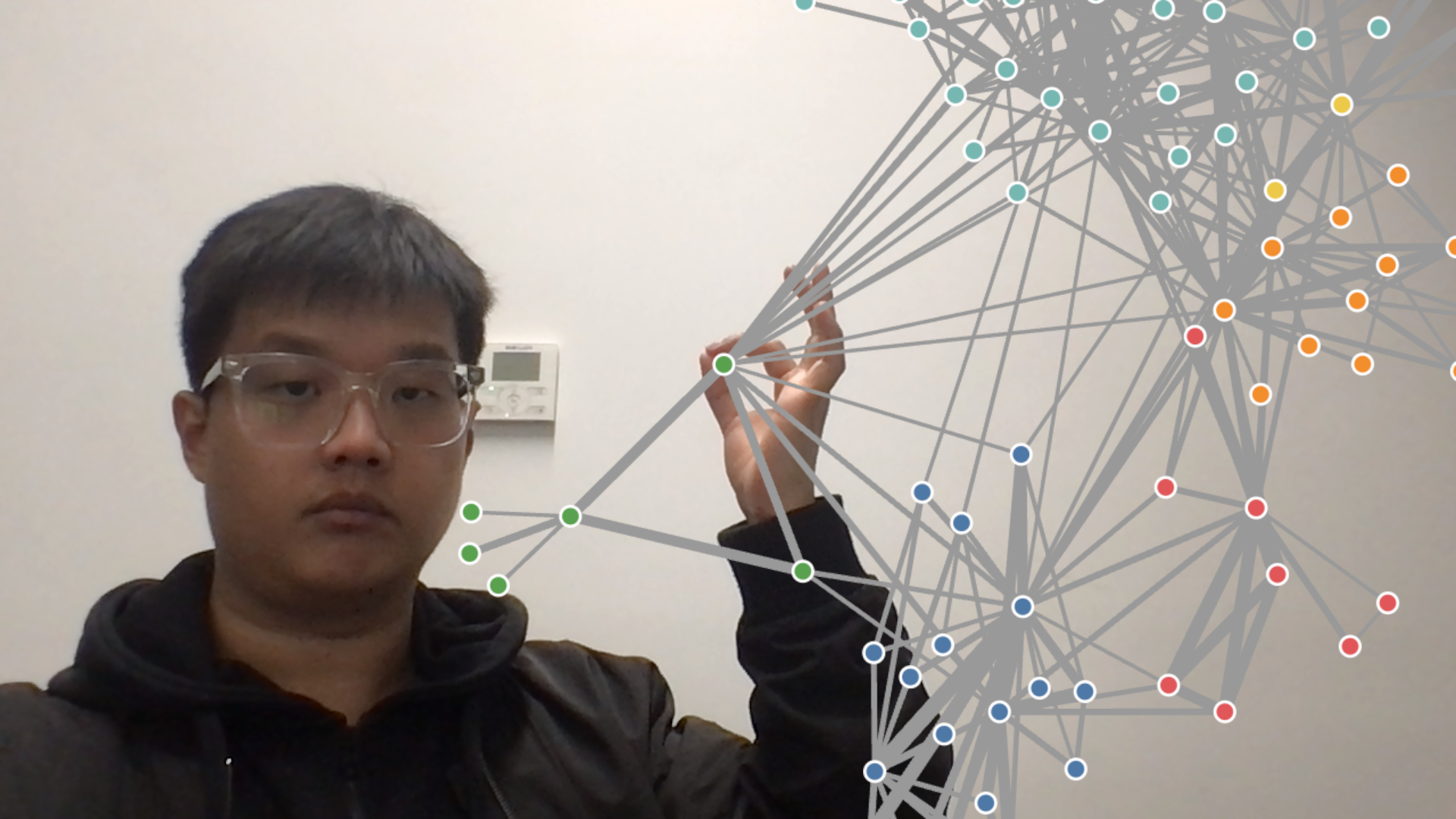}
  \caption{Live video-stream captures of our system. Left: we can use a ``selfie-segmentation'' model to place the data visualisation behind the presenter but in front of the background; middle and right: before and after dragging a node in a force-directed graph using a pinch and drag gesture.}
  \label{fig:teaser}
}

\abstract{
We present \textit{Hanstreamer}, a free and open-source system for webcam-based data presentation.
The system performs real-time gesture recognition on the user's webcam video stream to provide interactive data visuals. Apart from standard chart and map visuals, Hanstreamer is the first such video data presentation system to support network visualisation and interactive DimpVis-style time-series data exploration.
The system is ready for use with popular online meeting software such as Zoom and Microsoft Teams.
} 

\CCScatlist{
    \CCScatTwelve{Human-centered computing}{Human computer interaction}{Interaction paradigms}{Mixed / augmented reality}
    \CCScatTwelve{Human-centered computing}{Ubiquitous and mobile computing}{Ubiquitous and mobile devices}{Interactive whiteboards}
}

\begin{document}



\maketitle
\section{Introduction} 
Since the beginning of the COVID-19 pandemic, a significant part of the workforce has shifted to remote,
online work. When collaborating with their peers, workers share their screens to explain content,
communicate insights and make sense of data. In sectors such as research and business, data 
visualisations are often used to present study results or to report on key performance indicators. Typically, presentations over Microsoft Teams or Zoom consist of a full-screen view of the presenter's desktop application while the 
presenter's webcam feed is displayed in a small window in a corner. This style of on-line presentation emphasises the content and places the presenter in the background.
By contrast, we were inspired by data presentations by the late Hans Rosling \cite{hansroslingvideo}, who demonstrated that a view of the presenter with a clear, physical connection to the data visualisation is essential for engaging data storytelling and communication, but this is lost in conventional online meeting presentation formats.

In this paper, we present the design of a web-cam-based, interactive data visualisation tool that puts the 
presenter \emph{inside} the visualisation space (see Figure \ref{fig:teaser}). It is designed to facilitate casual data discussion
and promote audience engagement with the visualisations. Using image-based machine learning facilities
and web-based visualisation, it tracks a set of hand gestures and supports interactions with the 
visualisation such as details on demand, zooming and panning.  We designed a tool that allows for 
data presentations but also casual exploration and collaborative sense-making. 

Recently, Hall et al. \cite{chironomia2022uist} presented a system that allows a presenter to perform touchless interactions with hand gestures on visualisations in a remote meeting scenario.
It supports a variety of gesture-supported interactions, such as revealing the details of a data point in a chart by pointing with the index finger and re-positioning chart elements by pinching and dragging.
Furthermore, it features an assortment of basic charts, which are often used in many data presentations, namely bar, pie, area and line charts.
Compared to the system presented by Hall et al.~\cite{chironomia2022uist}, we provide:
\begin{itemize}
 \item a completely open-source system for integrating information graphics and gesture control into a video stream that can be shared with others in online meeting or presentation platforms, such as Zoom and Microsoft Teams;
 \item the first demonstration of interactive network visualisation manipulation in a live video stream data presentation; and
 \item an implementation of interactive time series navigation using gestures based on the DimpVis data-contextual time-scrubbing approach \cite{kondo2014dimpvis}.
\end{itemize}
\begin{figure}
    \centering
    \includegraphics[width=0.5\textwidth]{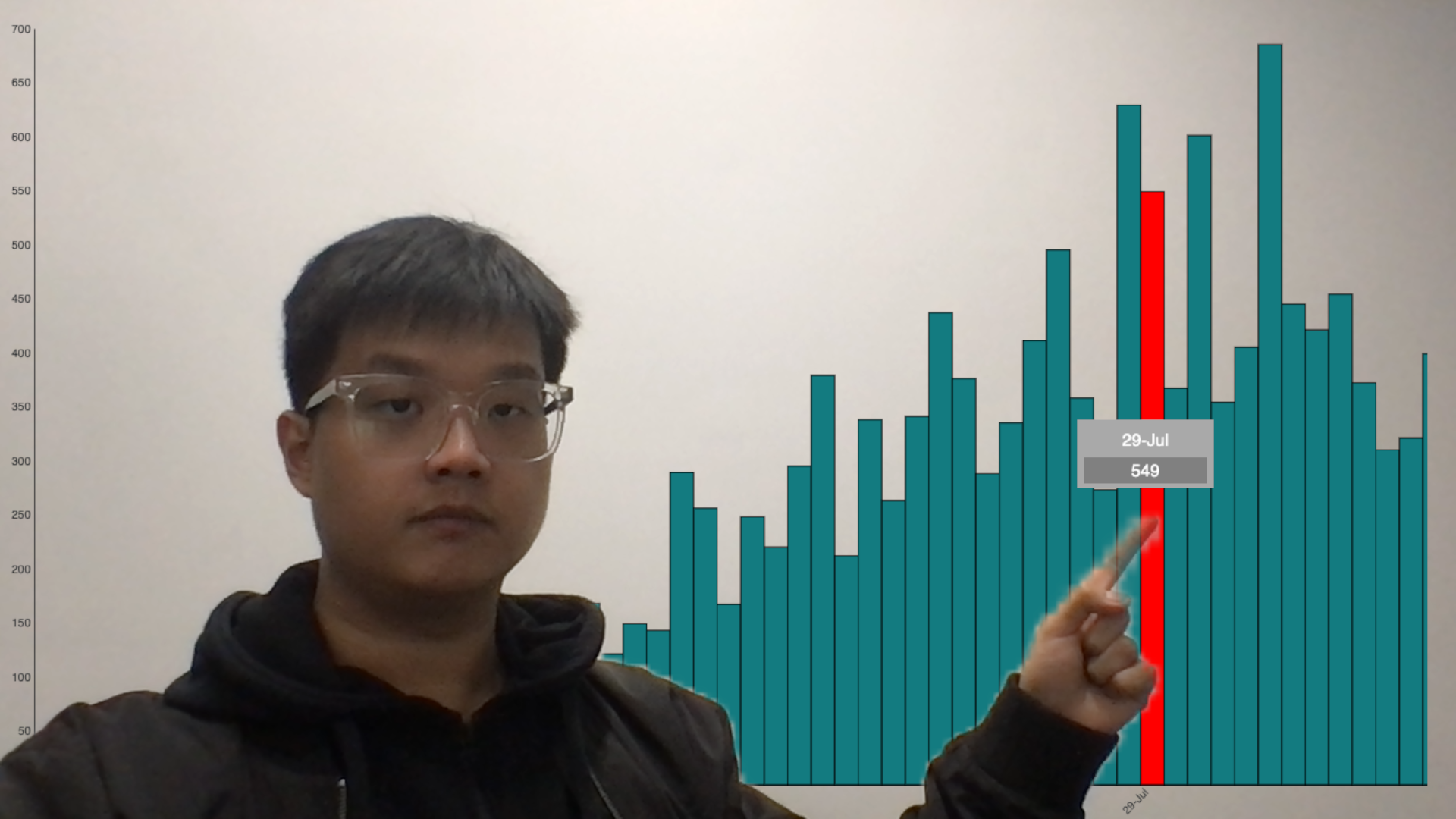}
    \caption{An example of the pointing gesture. The presenter points at a data point on the bar chart, revealing additional details.}
    \label{fig:pointing}
\end{figure}
\section{Goals and System Design}
The main goals of this work are to provide an open-source platform to foster research in the topic of data presentation with webcam-tracked gestures and to explore the design of the workflow and interactions needed for a presenter to prepare their story.

\begin{figure}
    \centering
    \includegraphics[width=0.5\textwidth]{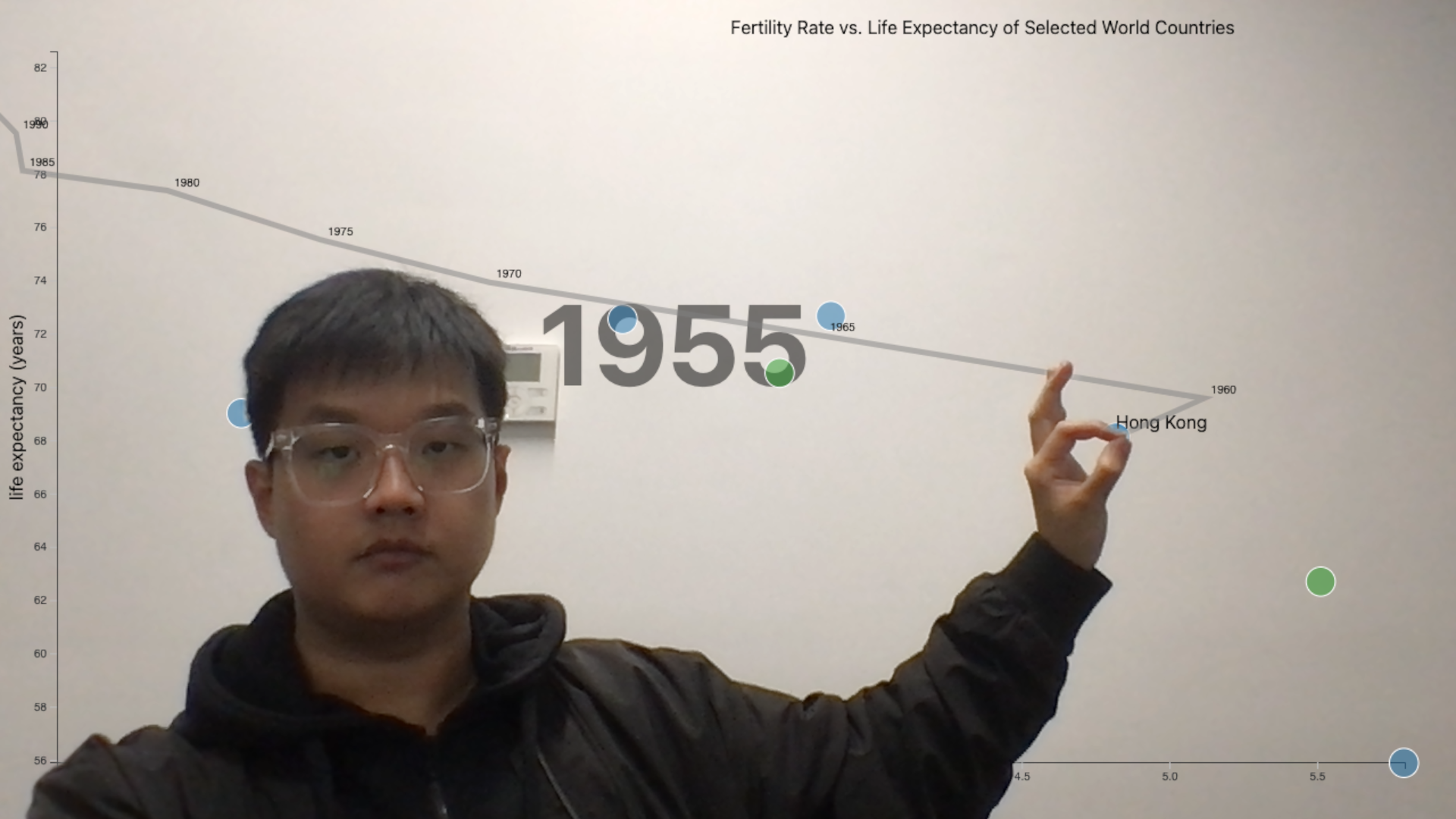}
    \includegraphics[width=0.5\textwidth]{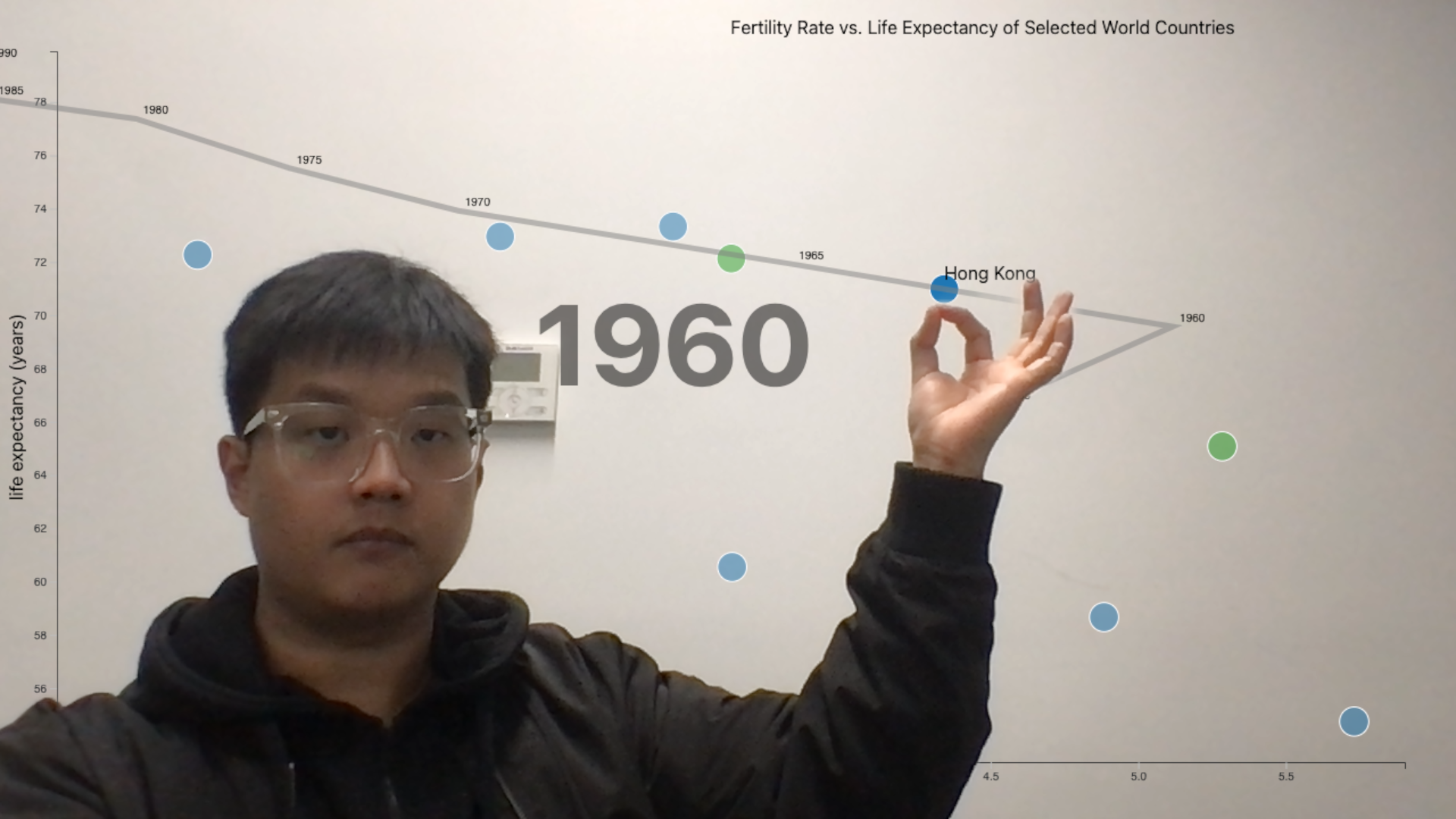}
    \caption{An example of the pinching gesture. In the DimpVis visualisation, the presenter pinches one of the data points and drags it along the pre-defined trajectory for that point, advancing the time step shown (i.e., from 1955 to 1960 in this example).}
    \label{fig:pinching}
\end{figure}
\subsection{An open-source platform for webcam-based data presentation}
The platform is currently being open-sourced at \href{https://github.com/adriankristanto/Hanstreamer}{https://github.com/adriankristanto/Hanstreamer}.
In addition to it being open-sourced, we decided to implement the platform as a web application, as it aligns with our goal to make this platform accessible to a wide range of users without being limited to specific operating systems.

The user interface components of the application are built with React due to its fast performance, while D3.js \cite{bostockDataDrivenDocuments2011} is utilised as the visualisation engine, such as for processing the data that is being visualised and controlling the zoom level of a chart, and renderer (in this case, the visualisations are rendered by D3.js on top of an HTML canvas element). In its current version, the platform supports several simple charts, including bar and multi-series line charts, alongside advanced visualisations, including network graph and DimpVis \cite{kondo2014dimpvis}.

As the use of body language is paramount during a presentation, we decided to include a feature that allows the presenter to put themselves in front of the visualisation to avoid obstructing the non-verbal communication aspect of the presentation. We use Mediapipe \cite{lugaresiMediaPipeFrameworkBuilding2019} selfie segmentation model to separate the presenter from the background. Then, the segmented portrait of the presenter is rendered on a new canvas layer, which is drawn over the webcam video stream. This allows the visualisation to be rendered between the webcam video stream and the topmost layer where the presenter is rendered.

\begin{figure}
    \centering
    \includegraphics[width=0.5\textwidth]{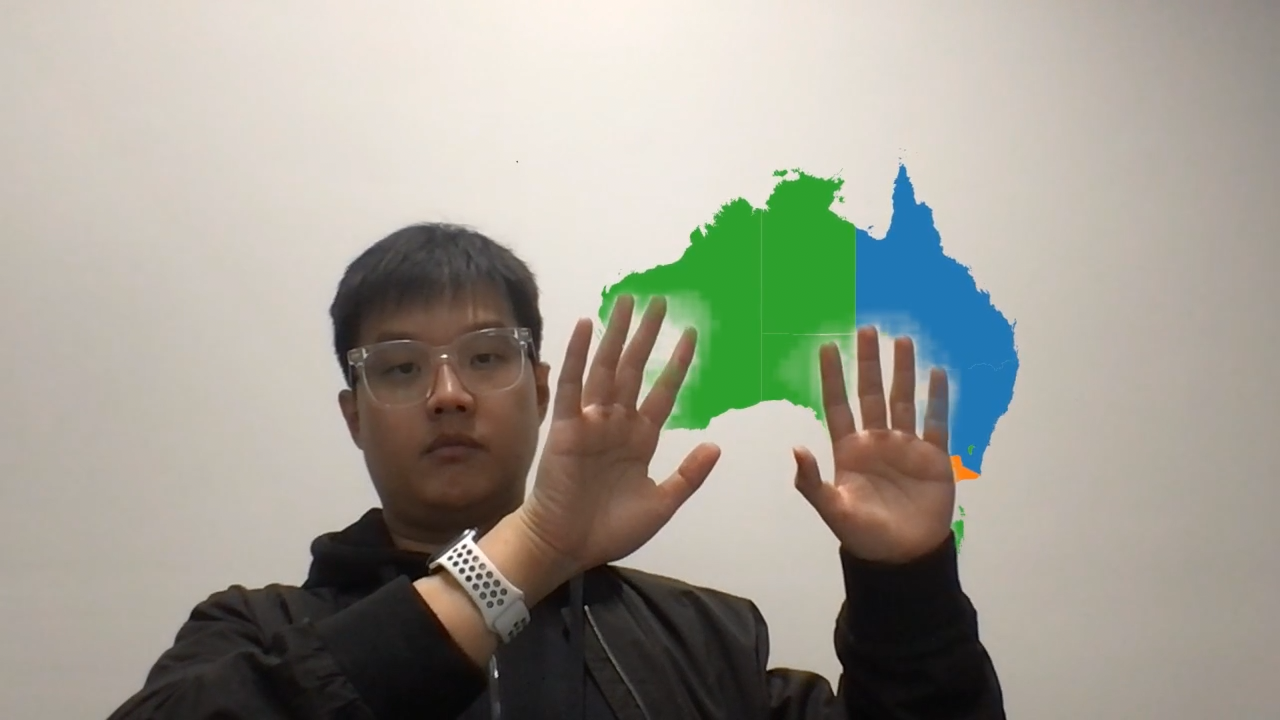}
    \includegraphics[width=0.5\textwidth]{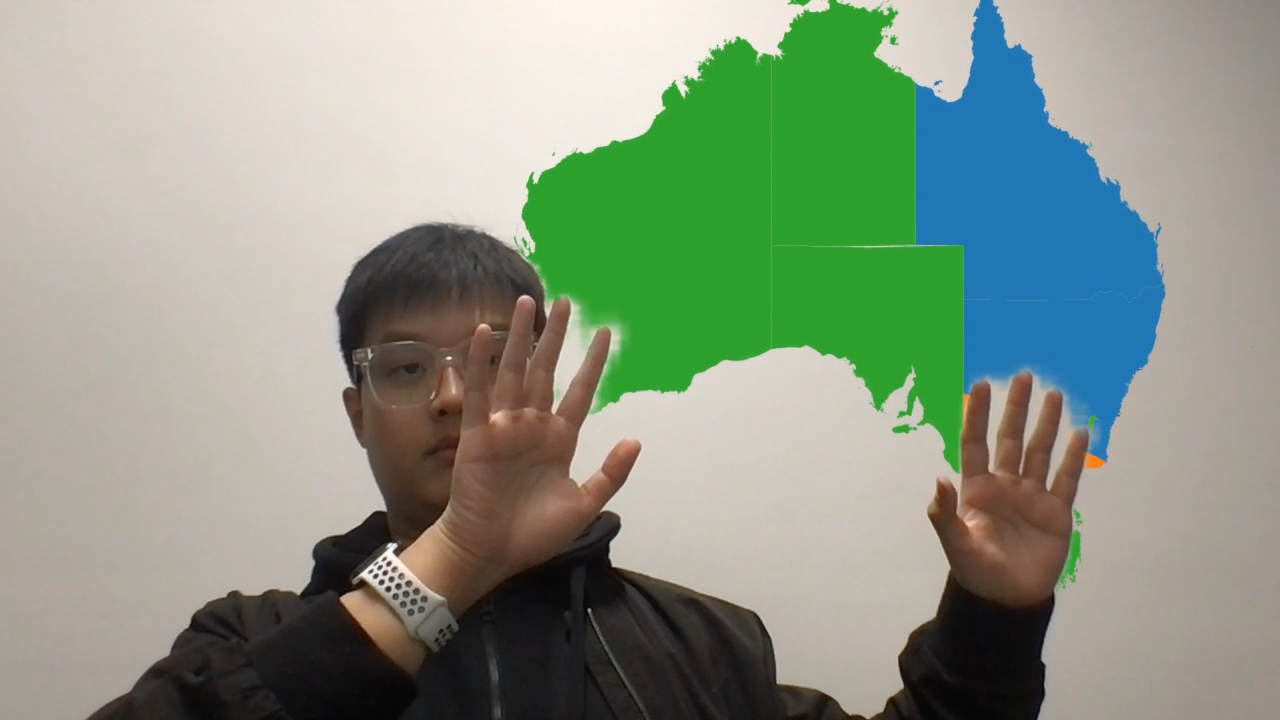}
    \caption{An example of the zooming gesture. The presenter moves their open palms away from each other to zoom in on the chart.}
    \label{fig:zooming}
\end{figure}

A core feature of the platform is the set of gestures that allow the presenter to interact with the charts directly in the visualisation space. This feature is implemented as a two-step process: hand detection and gesture recognition. Firstly, the platform incorporates Mediapipe’s hand landmark detection model to detect the presenter’s hands in the webcam video stream. Next, instead of using yet another machine learning model to recognise the hand gestures of the presenter, which could potentially deteriorate the performance of the application, we used a slightly modified version of fingerpose.js\footnote{\href{https://github.com/andypotato/fingerpose}{https://github.com/andypotato/fingerpose}} that uses the cosine rule to determine how curled each finger is to determine the current hand gesture of the presenter. At the time of development, a modification was required to support recognising gestures for both hands instead of a single hand. This two-step process of hand detection and gesture recognition allows us to support a variety of hand gestures, including:

\begin{figure}
    \centering
    \includegraphics[width=0.5\textwidth]{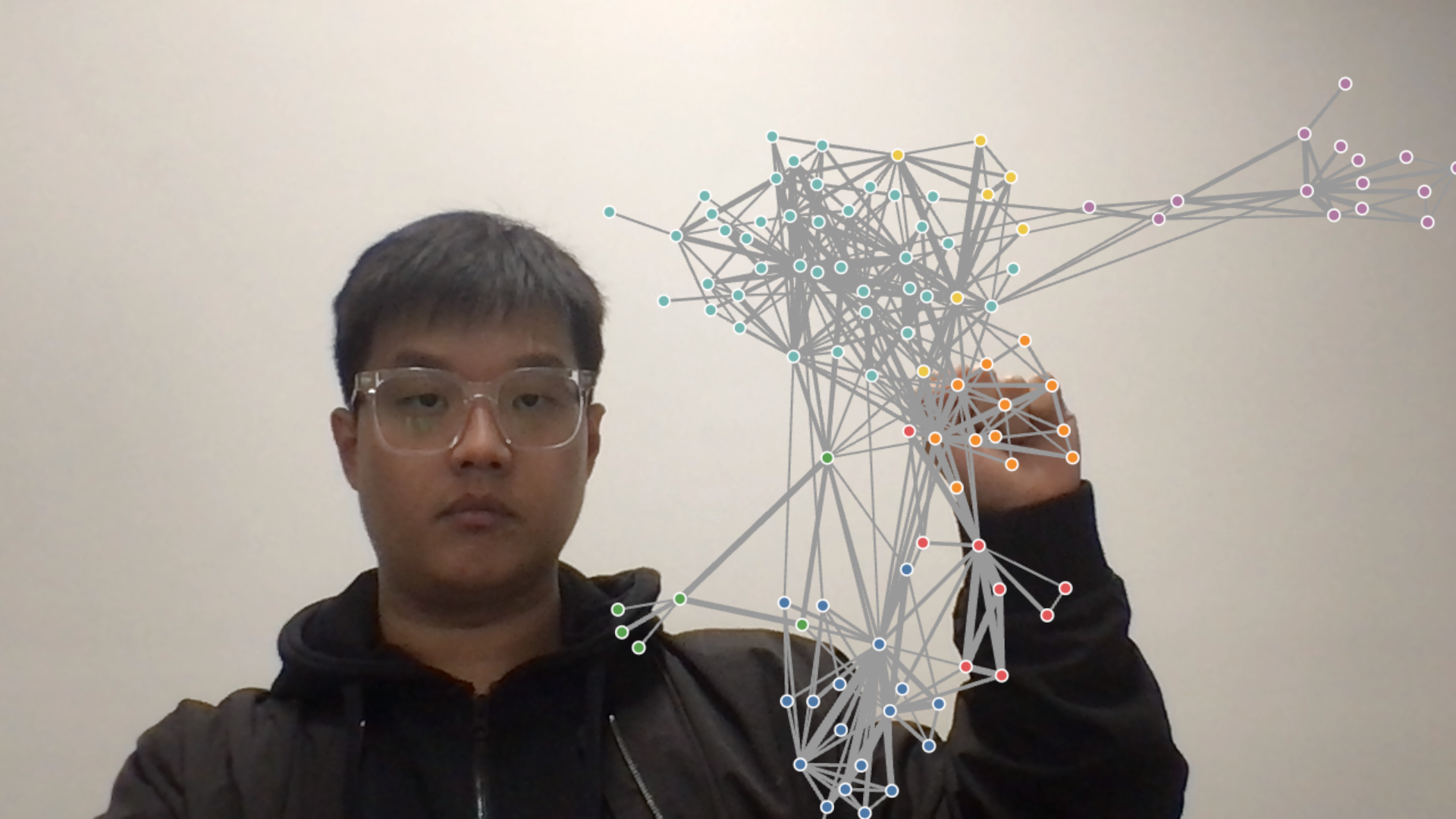}
    \includegraphics[width=0.5\textwidth]{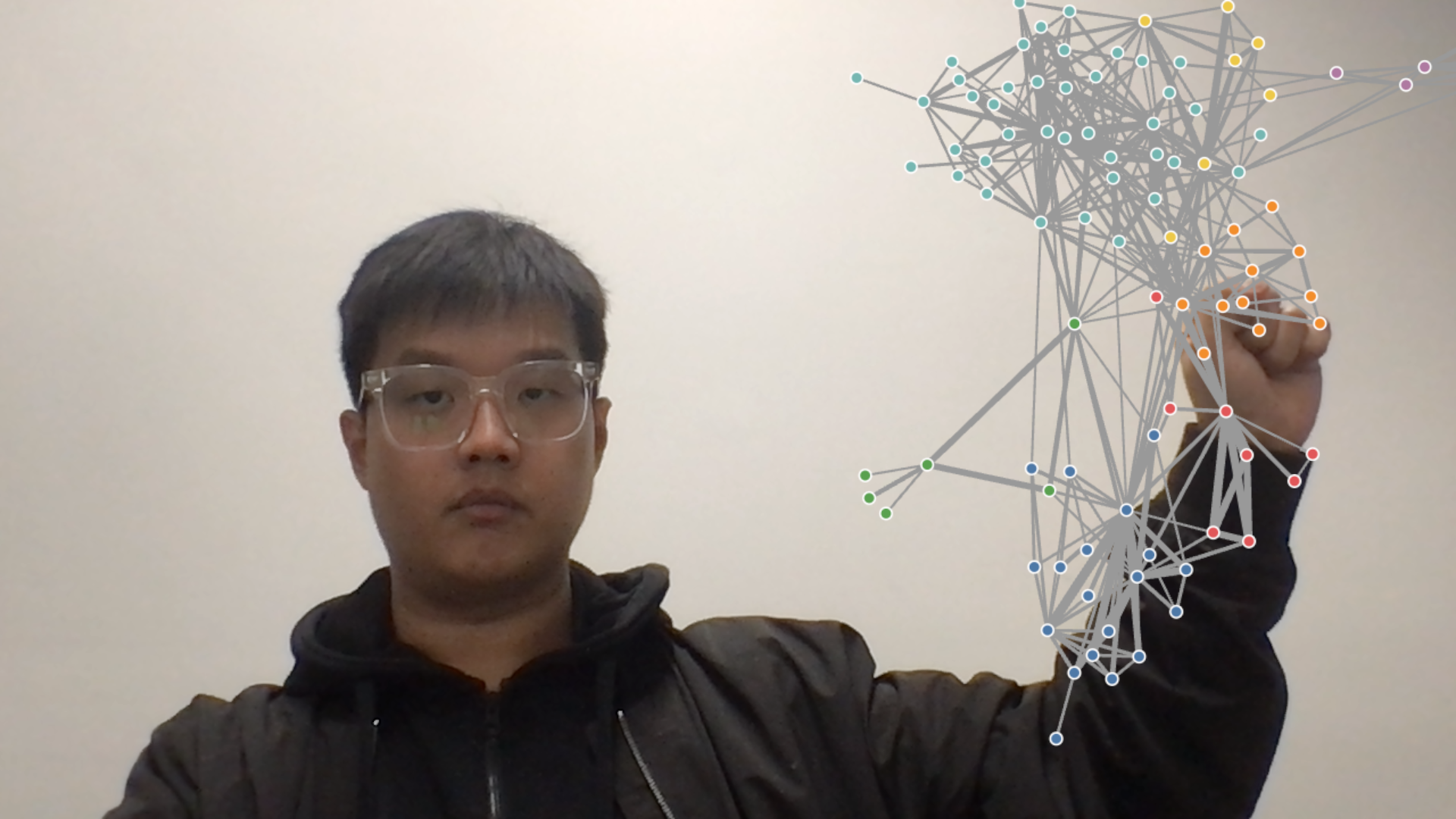}
    \caption{An example of the panning gesture. The presenter ``grabs`` the visualisation and moves it freely within the visualisation space.}
    \label{fig:panning}
\end{figure}
\begin{itemize}
    \item Pointing (see Figure \ref{fig:pointing}): This gesture allows the presenter to highlight a certain data point in the visualisation by pointing their index finger to the data point. In the current implementation, this gesture is supported by all charts.
    \item Pinching (see Figure \ref{fig:pinching}): The presenter can perform this gesture by connecting the thumb and index finger to form a circle, while the other fingers are kept straight on a specific data point. This gesture is supported in the network graph, where the user can pinch and pull one data point in the graph, and DimpVis, where the user can drag one data point following a specific path to see how the graph progresses over time. 
    \item Panning (see Figure \ref{fig:panning}): This gesture can be performed by making a fist on top of the visualisation. In the current version, this gesture is supported by all visualisations.
    \item Zooming (see Figure \ref{fig:zooming}): The presenter opens both of their palms and moves them away from each other to zoom in on the visualisation. To zoom out, the presenter can move their palms in the opposite direction.
\end{itemize}

This application can also be used directly as a camera source on online meeting platforms, such as Zoom and Microsoft Teams, with the use of OBS Studio\footnote{\href{https://obsproject.com/}{https://obsproject.com/}}. The instruction on how to configure the tools to work with OBS is available at \href{https://github.com/adriankristanto/Hanstreamer}{https://github.com/adriankristanto/Hanstreamer}.

\subsection{An exploration of storytelling preparation interface}
This work also addresses the need for the presenter to prepare their stories, i.e., which visualisations need to be presented, in which sequence, which interaction is needed to perform transitions and which interactions.

To address these questions we provide an initial design of a \emph{story planner}, a visual interface that lets the presenter prepare the sequence of visualisations, transitions and associated interactions. Visualisations and interactions are represented by stacked, green vertical bars that can be rearranged in the sequence using drag and drop. The user can define transition and interaction gestures, as described in the previous section.

At this stage, the workflow is not fully dynamic; the user prepares the sequence and follows a scripted visualisation, which dictates the visualisation order and the type of interactions. We would like to explore more dynamic approaches to this by extending our gesture set to include gestures that allow the user to dynamically navigate the story.  For example, to switch arbitrarily to other visualisations in the sequence by swiping the visualisations, or via a menu.





\label{sec:systemdesign}

\section{Conclusion and Future Work}
We propose an open-source toolkit to perform live presentations with data visualisations. The initial stage of the toolkit includes design choices on interactions with the visualisations and on the way that presenters create their stories. Several research questions need to be addressed to further understand how this platform can enhance data presentations in remote meetings. In particular, we would like to address:
\begin{itemize}
    \item ``What level of complex interaction is required to present advanced visualisation stories?''
    \item ``What is the design space of a dynamic data visualisation presentation tool?''
\end{itemize}  
We plan to explore these questions through iteratively developing prototype features in Hanstreamer and evaluating their efficacy via user testing.

\acknowledgments{
}

\bibliographystyle{abbrv-doi}

\bibliography{refereces}

\end{document}